\documentclass[reprint,prl,superscriptaddress]{revtex4-1}
\usepackage{graphicx}
\usepackage[utf8]{inputenc}
\usepackage{latexsym}
\usepackage{amsmath}
\usepackage{amssymb}
\usepackage{bm}
\usepackage{float}
\usepackage{hyperref}
\usepackage[outercaption]{sidecap}
\usepackage{url}
\usepackage{etoolbox}
\usepackage{xcolor}

\newcommand{\ket}[1]{| #1\rangle}

\newcommand{\hide}[1]{}

\begin{document}

\title{$T^3$-Stern-Gerlach Matter-Wave Interferometer}

\author{O. Amit}
\thanks{Email address: amio@bgu.ac.il}
\address{Department of Physics, Ben-Gurion University of the Negev, Be'er Sheva 84105, Israel}
\author{Y. Margalit}
\address{Department of Physics, Ben-Gurion University of the Negev, Be'er Sheva 84105, Israel}
\address{Research Laboratory of Electronics, MIT-Harvard Center for Ultracold Atoms, Department of Physics,
Massachusetts Institute of Technology, Cambridge, Massachusetts 02139, USA}
\author{O. Dobkowski}
\author{Z. Zhou}
\author{Y. Japha}
\address{Department of Physics, Ben-Gurion University of the Negev, Be'er Sheva 84105, Israel}
\author{M. Zimmermann}
\author{M. A. Efremov}
\address{Institut f\"ur Quantenphysik and Center for Integrated Quantum Science and Technology ($IQ^{ST}$), Universit\"at Ulm, D-89081 Ulm, Germany}
\author{F. A. Narducci}
\address{Department of Physics, Naval Postgraduate School, Monterey, CA 93943, USA}
\author{E. M. Rasel}
\address{Institut f\"ur Quantenoptik, Leibniz Universit\"at Hannover, D-30167 Hannover, Germany }
\author{W. P. Schleich}
\address{Institut f\"ur Quantenphysik and Center for Integrated Quantum Science and Technology ($IQ^{ST}$), Universit\"at Ulm, D-89081 Ulm, Germany}
\affiliation{Hagler Institute for Advanced Study at Texas A$\&$M University,
Texas A$\&$M AgriLife Research, Institute for Quantum Science and Engineering (IQSE), and Department of Physics and Astronomy,
Texas A$\&$M University, College Station, TX 77843-4242, USA}
\affiliation{Institute of Quantum Technologies, German Aerospace Center (DLR), D-89069 Ulm, Germany}
\author{R. Folman}
\address{Department of Physics, Ben-Gurion University of the Negev, Be'er Sheva 84105, Israel}

\begin{abstract}

We present a unique matter-wave interferometer whose phase scales with the cube of the time the atom spends in the interferometer.
Our scheme is based on a full-loop Stern-Gerlach interferometer incorporating four magnetic field gradient pulses to create a
state-dependent force.
In contrast to typical atom interferometers which make use of laser light for the splitting and recombination of the wave packets, this
realization uses no light and can therefore serve as a high-precision surface probe at very close distances.

\end{abstract}

\maketitle

\noindent{\it Introduction.--}
The Stern-Gerlach (SG) effect \cite{Gerlach1922} of 1922 is a paradigm of quantum mechanics and allows an illuminating  glimpse
\cite{Schwinger2001} into the inner workings of this theory. Moreover, it arguably marks the birth of atom interferometry
\cite{Berman1997,Cronin2009}. Indeed, the splitting of an atomic beam by a magnetic field gradient served as the starting point for David Bohm
\cite{Bohm1951} and Eugene Paul Wigner \cite{Wigner1963} in their discussion of the coherence in a SG interferometer (SGI). A detailed
analysis \cite{Englert1988, Schwinger1988, Scully1989} of such a device concluded that it would require an extreme accuracy of the field
gradients to maintain coherence, a formidable challenge appropriately coined the Humpty Dumpty effect.

In the present Letter we report on the successful implementation of an SGI utilizing the strong and accurate magnetic field gradients
\cite{Machluf2013,Margalit2018} provided by the currents in the wires of an atom chip \cite{Keil2016}. Our SGI is unique in three aspects: (i)
Although the gradient fields act on the atom continuously during its flight through the interferometer, as in the Humpty-Dumpty configuration
\cite{Englert1988,Schwinger1988,Scully1989}, we obtain a remarkably high contrast. (ii) The observed phase shift scales \cite{Zimmermann2017}
with the cube of the time the atom spends in the SGI, and thus represents the first interferometric measurement of the Kennard phase
\cite{Kennard1927,Kennard1929,Rozemann2019} predicted in 1927. (iii) The lack of light pulses to split and recombine the beams in combination
with the Kennard phase  makes our interferometer a perfect probe for magnetic as well as other properties of surfaces.

In general, the phase of spatial light-pulse atom interferometers \cite{Berman1997,Cronin2009} does not display a pure $T^3$-scaling, where
$T$ is the total interferometer time. For example, in the Ramsey-Bord\'e interferometer \cite{BORDE1989}, the phase shift originates from a
constant {\it position} difference between two paths, which in the presence of a time-independent linear potential leads to a phase
proportional to $T$. In the Kasevich-Chu interferometer \cite{Kasevich1991,Peters2001}, it is caused by a piece-wise constant {\it momentum}
difference leading to a $T^2$-scaling. However, our $T^3$-SGI has a piece-wise constant {\it acceleration} difference between its two paths
resulting in a phase scaling as $T^3$.

Phase contributions scaling with $T^3$ emerge also in other setups and originate, for example, from rotation in a Sagnac interferometer
\cite{Dutta2016}, the presence of a gravity gradient \cite{Kasevich2017} or a time-dependent acceleration induced by Bloch oscillations
\cite{McDonald2014} in the Kasevich-Chu configuration. However, our $T^3$-SGI has a \textit{pure} $T^3$-scaling resulting from the imprinted
{\it state-dependent linear potentials}.

Probing surfaces with sensitive matter-waves is a long standing goal. Previous studies have probed short-distance phenomena such as
Casimir-Polder forces \cite{Harber2005,Slama2010}, Johnson noise \cite{Jones2003,Blatt2015}, patch potentials
\cite{Blatt2015,Chan2014,Lakhmanskiy2019}, and exotic physics such as the fifth force \cite{Sorrentino2009,Santos2018,comment-Santos}. Matter-wave
interferometry near the surface can significantly increase the sensitivity. Unlike a light-based atom interferometer, which cannot operate
close to the surface due to light scattering from the nearby object, the SGI does not require any laser light in order to control a coherent
spatial superposition of two atomic wave packets. As the accumulated
phase is sensitive to magnetic fields, the SGI may be used as a unique probe for magnetic surface properties,
as well as for noise and order parameters in electron transport such as squeezed currents. In addition, in a
``matter-wave homodyne'' type scheme, one wave packet may be put in the vicinity of the surface while the other
acts as a reference.

\noindent{\it Setup.--} Our experiment is based on what is, to the best of our knowledge, the first full-loop SGI realization
\cite{Margalit2018}, as originally envisioned \cite{Bohm1951,Wigner1963}. Different from previous realizations of the SG effect
\cite{Gerlach-optics1,Gerlach-optics2}, it uses four magnetic field gradient operations for splitting, stopping, accelerating back, and
stopping for a complete closure of a full-loop. The experiment begins with a $^{87}$Rb Bose-Einstein condensate (BEC) released from a magnetic
trap and falling freely under gravity. We define our coordinate system such that the $z$-axis is along the gravitational acceleration of magnitude $g$.
The non-linear Zeeman effect of a bias magnetic field $B_0$ of $35$\,G along the $y$-axis, added by an external pair of Helmholtz coils,
creates an effective two-level system consisting of the states $\ket{1}\equiv\ket{F=2,m_{F}=1}$ and $\ket{2}\equiv\ket{F=2,m_{F}=2}$ of the
$5^2 S_{1/2}$ manifold. Since the BEC quickly expands after its release from the magnetic trap, the interatomic interactions are negligible
and the experiment may be considered a single-particle experiment.

\begin{figure}
\centerline{
\includegraphics[trim={3.4cm 1.8cm 5.2cm 1.7cm},clip,height=0.65\columnwidth,width=\columnwidth]{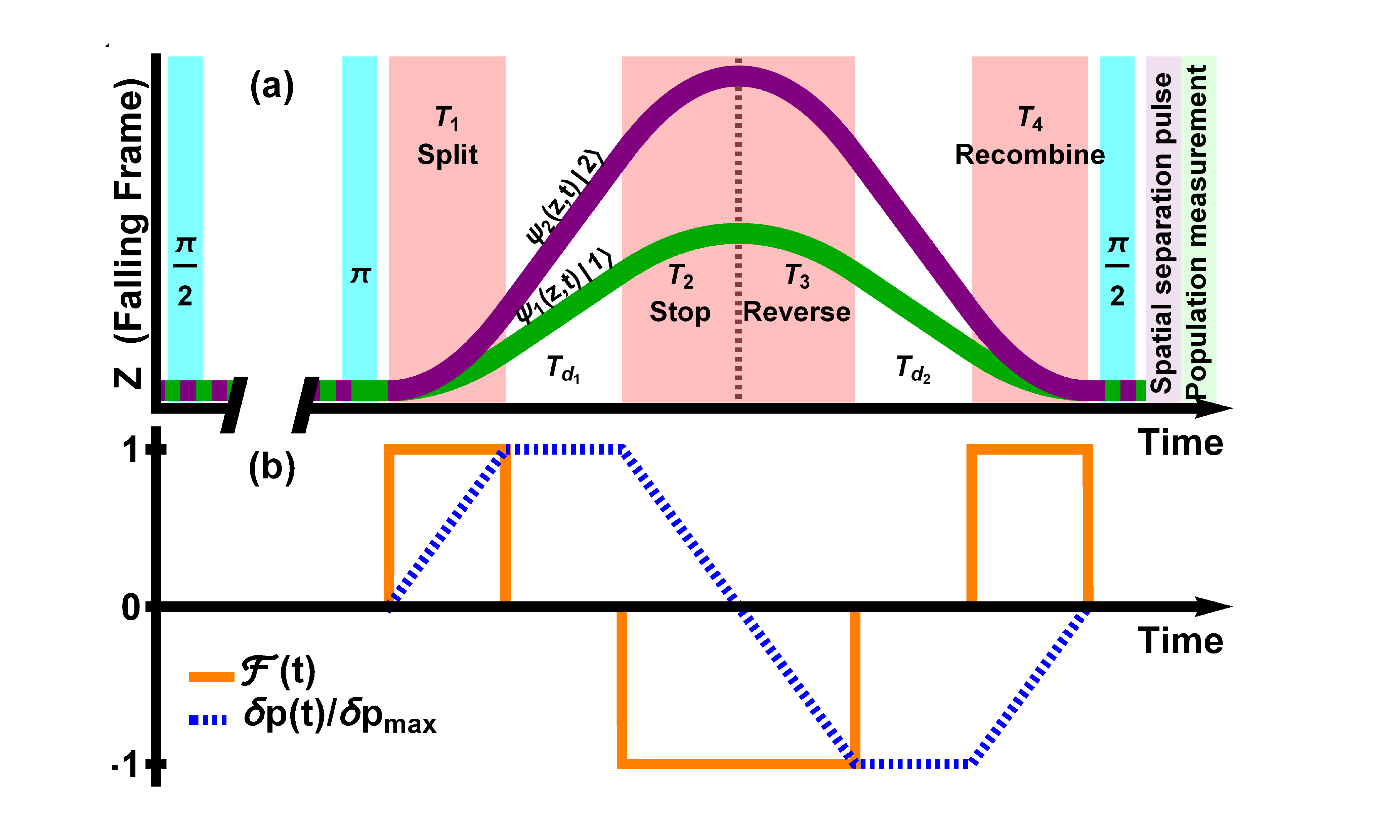}}
\caption{Pulse sequence of our longitudinal $T^3$-SGI (not to scale).
(a) Center-of-mass trajectories of the atomic wave packets with internal states $\ket{1}$ (bottom, green) and $\ket{2}$ (top, purple). Also
shown are the $\pi/2$ and $\pi$ RF (light blue) and magnetic gradient splitting, stopping, reversing and recombining pulses (light red). The
magnetic field gradients result in a state-dependent force along the z-direction while the strong bias magnetic field along the y-direction
defines the quantization axis.
(b) Time-dependence of the relative force ${\mathcal F}\equiv{\mathcal F}(t)$, Eq.~(\ref{F_dif_t}), (orange curve) and the corresponding
relative momentum $\delta p(t)=\mathcal{P}_2(t)-\mathcal{P}_1(t)$, (blue dashed curve) between the wave packets moving along the two
interferometer paths. In the experiment, we have achieved the maximal separations
$\delta z_{\rm{max}}=1.4$ $\mu$m and $\delta p_{\rm {max}}/m_{Rb}=19$ mm/s in position and velocity.
}
\label{fig:seq}
\end{figure}

After being released, the initially prepared atomic internal state $\ket{2}$ is transferred to an equal superposition,
$\ket{\phi_{\pi/2}}\equiv (\ket{1}+\ket{2})/\sqrt{2}$, by an on-resonance radio-frequency (RF) $\pi/2$-pulse, as shown in Fig.\,\ref{fig:seq}.
Following a free-fall time of $400\,\mu$s, named hereafter the dark time, we apply an RF $\pi$-pulse that flips the atomic state to
$(\ket{1}-\ket{2})/\sqrt{2}$. After a second dark time of another $400\,\mu$s, the second RF $\pi/2$-pulse completes the echo sequence. Using
a long magnetic field gradient and standard absorption imaging, we measure the relative populations of the states.

The interferometric scheme realized during the second dark time consists of four magnetic field gradient pulses produced by three parallel
gold wires along the $x$-axis on the atom chip \cite{Margalit2018}. The currents running through them in alternating directions are identical
and create a two-dimensional quadrupole field in the $yz$ plane with its center (zero) at $z_0=98\,\mu$m below the chip surface.

The first gradient pulse, applied for a duration of $T_1$, prepares the state of the atomic center-of-mass motion in the form of two wave packets taking separate trajectories. Indeed, close to the center, the wave packets experience different forces ${\bf F}_{i}=\mu_{i}(\partial B_y/\partial z){\bf e}_z$ where $\mu_i\equiv \langle i|{\hat{\mu}_y}| i\rangle= \mu_B g_{F_i}m_{F_i}$ is the mean value of the magnetic dipole moment of the state $\ket{i}$ with $i=1,2$. Here $\mu_B$, $g_F$, and $m_F$ denote the Bohr magneton, the Land\'e factor, and the projection of the angular momentum on the $y$-axis, respectively.

After a delay time, $T_{d_1}=T_d$, which is limited by the speed of our electronic circuits, we apply a second and third pulse back-to-back of
identical duration, $T_{2}=T_{3}=T_1$. This combined pulse, having opposite polarity (the direction of the magnetic field gradient) to the
first pulse thereby applying a force in the opposite direction, first compensates the momentum difference between the two wave packets, and
then reverses the direction of their motion. A fourth pulse applied after another delay time, $T_{d_2}=T_d$, for a duration $T_4=T_1$, with
the same polarity as the first pulse, leads to an interferometer which is closed both in momentum {\it and} position.

\begin{figure*}
\centerline{
\includegraphics[height=0.3\textwidth,width=\textwidth]{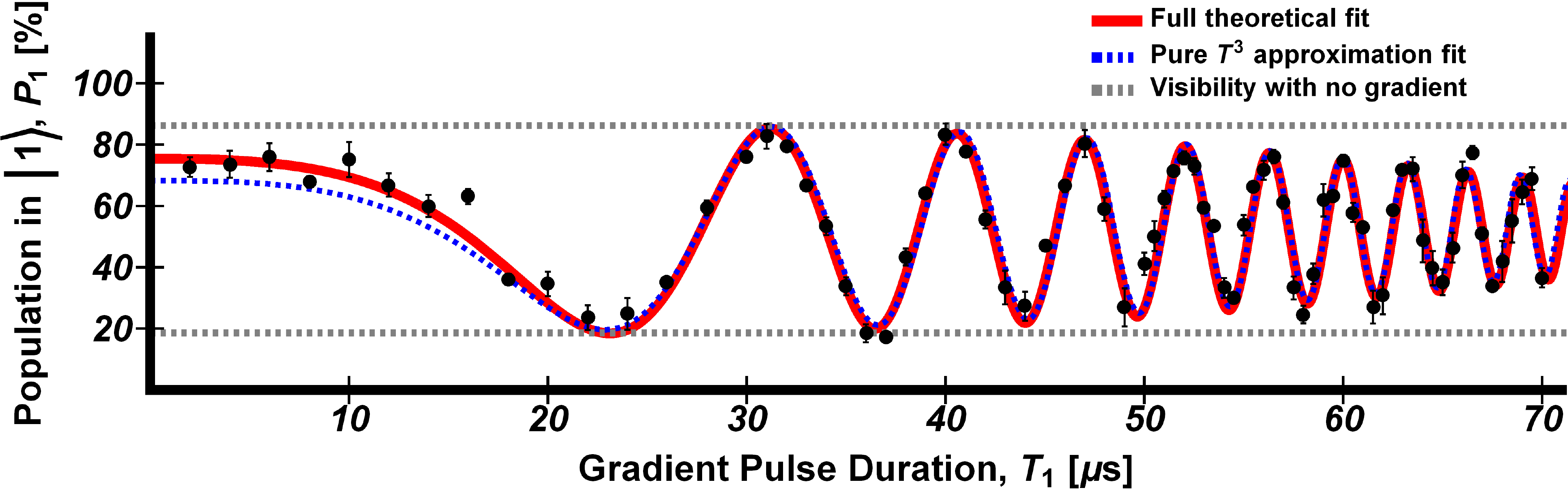}}
\caption{Measurement of the cubic phase with the $T^3$-SGI presented in Fig.~\ref{fig:seq}.
The solid red line represents a fit based on Eqs. (\ref{P1}) and (\ref{phase_result}) to the experimental data (dots), where the fitting
parameters are the decay constant, the magnetic acceleration $a_B$, and a constant phase $\varphi_0$. The dashed blue line is a fit based on
Eq.\,(\ref{phase_result}) with $T_{d}=0$, leading to a pure $T_1^3$ scaling of the interferometer phase as seen in
Eq.\,(\ref{phase_result_1}). The visibility drops from $68\%$ to $32\%$ over $70\,\mu {\rm s}$ with a decay time of $75\,\mu {\rm s}$. This
reduction results from inaccuracies in recombining the two interferometer paths. The dashed gray horizontal lines depict the maximal and
minimal values of the population $P_1$ measured without magnetic field gradients.}
\label{fig:closedloop}
\end{figure*}

\noindent{\it Phase of interferometer.--} In the $T^3$-SGI,
no transitions occur during the four gradient pulses, as shown in Fig.\,\ref{fig:seq}.
Hence, the complete quantum state
\begin{equation}
 \label{state tf}
  \ket{\Psi(t)}=\frac{1}{\sqrt{2}}\left[\ket{1}\hat{U}_{1}(t,0)\ket{\psi(0)}-
  \ket{2}\hat{U}_{2}(t,0)\ket{\psi(0)}\right]
\end{equation}
is determined by the state $\ket{\psi(0)}$ of the center-of-mass motion directly after the $\pi$-pulse, with $\langle \psi(0)|\psi(0)\rangle =1$,
together with the time evolution operator $\hat{U}_{i}$ for the internal state $\ket{i}$ corresponding to the time-dependent potential
\begin{equation}
 \label{potential V}
  V_{i}(z,t)\equiv -\left[mg+ \mu_i \frac{\partial B_y}{\partial z}\mathcal{F}(t)\right]z
\end{equation}
where $m$ is the atomic mass.

Moreover, the time dependence
$$
{\mathcal F}(t)\equiv\Theta(t)-\Theta(t-T_1)-\Theta(t-T_1-T_d)+\Theta(t-3T_1-T_d)
$$
\begin{equation}
 \label{F_dif_t}
  +\Theta(t-3T_1-2T_d)-\Theta(t-4T_1-2T_d),
\end{equation}
displayed in Fig.~\ref{fig:seq}(b) by the orange curve involves the Heaviside step function $\Theta(t)$.

Since $V_i$ is linear in the coordinate $z$, we can obtain the explicit expression \cite{Zimmermann2019}
\begin{equation}
 \label{U_F}
  \hat{U}_i(t,0)\equiv e^{i\Phi_i(t)}\hat{D}\left[\mathcal{Z}_i(t),\mathcal{P}_i(t)\right]\hat{U}_{0}(t,0)
\end{equation}
for $\hat{U}_i$ which consists of the product of (i) the time evolution operator $\hat{U}_{0}(t,0)\equiv
\exp\left[-it\hat{p}_z^2/(2m\hbar)\right]$ of the center-of-mass motion of the free atom,
(ii) the displacement operator $\hat{D}\left[\mathcal{Z},\mathcal{P}\right]\equiv
\exp\left[(i/\hbar)\left(\mathcal{P}\hat{z}-\mathcal{Z}\hat{p}_z\right)\right]$ with
\begin{equation}
 \label{R and P}
  \mathcal{P}_i(t)\equiv -\int\limits_{0}^{t}d\tau \frac{\partial V_i}{\partial z}\;\;\;{\rm and}\;\;\;
  \mathcal{Z}_i(t)\equiv-\frac{1}{m}\int\limits_{0}^{t}d\tau (t-\tau)\frac{\partial V_i}{\partial z},
\end{equation}
as well as the position $\hat{z}$ and momentum $\hat{p}_z$ operators, and (iii) a phase factor
\begin{equation}
 \label{Phi_F}
  \Phi_i(t)\equiv -\frac{1}{2\hbar}\int\limits_{0}^{t}d\tau \mathcal{Z}_i(\tau)\frac{\partial V_i}{\partial z}.
\end{equation}

The second $\pi/2$-pulse applied at $t=T\equiv 4T_1+2T_d$ recombines both branches of the interferometer, and the probability
$P_1\equiv\langle \psi_1(T)|\psi_1(T)\rangle$ to observe atoms in the internal state $|1\rangle$
follows from the state $\ket{\psi_1(T)}\equiv \langle \phi_{\pi/2}|\Psi(T)\rangle$ of the atomic center-of-mass motion, which according to Eq.
(\ref{state tf}) takes the form
\begin{eqnarray}
 \label{phi 1}
  \ket{\psi_1(T)}& =&\frac{1}{2}\left\{e^{i\Phi_{1}(T)}\hat{D}\left[\mathcal{Z}_1(T),\mathcal{P}_1(T)\right]\right.\nonumber \\
  & -& \left. e^{i\Phi_{2}(T)}\hat{D}\left[\mathcal{Z}_2(T),\mathcal{P}_2(T)\right]\right\}\hat{U}_{0}(T,0)\ket{\psi(0)}.
\end{eqnarray}

From Eqs.\,(\ref{R and P}) and (\ref{F_dif_t}) we note that the two branches overlap
in both momentum and position, that is $\mathcal{P}_1(T)=\mathcal{P}_2(T)$ and $\mathcal{Z}_1(T)=\mathcal{Z}_2(T)$, as depicted in
Fig.\,\ref{fig:seq}, leading to identical displacement operators in Eq. (\ref{phi 1}). Hence, the probability
\begin{equation}
 \label{P1}
 P_1=\frac{1}{2}\left[1-\cos\left(\delta\Phi+\varphi_0\right)\right]
\end{equation}
contains the interferometer phase $\delta\Phi\equiv \Phi_{1}(T)-\Phi_{2}(T)$ given explicitly by
$$
\delta\Phi=\frac{mga_B}{\hbar}\left(\frac{\mu_1-\mu_2}{\mu_B}\right)\left(2T_1^3+3T_1^2T_d+T_1T_d^2\right)
$$
\begin{equation}
 \label{phase_result}
+\frac{ma_B^2}{\hbar}\left(\frac{\mu_1^2-\mu_2^2}{\mu_B^2}\right)\left(\frac{2}{3}T_1^3+T_1^2T_d\right).
\end{equation}
Here $\varphi_0$ is a constant phase taking into account possible technical misalignment and $a_B\equiv (\mu_B/m)(\partial B_y/\partial z)$.

In this derivation we have made two assumptions:
(i) Close to the center, the magnetic field generated by the three-wire configuration is linear, and (ii) the lengths of the four gradient
pulses are identical, that is $T_{2,3,4}=T_1$, and so are the two delay times, $T_{d_1}=T_{d_2}=T_d$.

Assumption (i) holds true when the distance ($\sim1\,\mu$m) traveled by the atomic wave packets is small compared to the distance
($\sim100\,\mu$m) from the chip. In our current apparatus, a violation of this constraint and the resulting magnetic field non-linearity gives
rise to a $3.5\%$ change in the applied force. Future improvements of the experiment will be required to address this issue.

Regarding assumption (ii), we have slightly adjusted the length of $T_4$, in order to better optimize the visibility and account for the
non-linearity, up to a difference of $8\%$ from $T_1$.

\noindent{\it Measurement of the cubic interferometer phase.--} Nevertheless, the experimental data (dots) depicted in
Fig.~\ref{fig:closedloop} agree very well with the theory (solid red line)
based on Eq.\,(\ref{phase_result}), where the fitting parameters are  the decay constant of the visibility, the magnetic field gradient
$\partial B_y/\partial z$ or $a_B$, and a constant phase $\varphi_0$. The dashed blue line is a fit based on Eq.\,(\ref{phase_result}) with
$T_{d}=0$, leading to a pure cubic scaling \cite{T3-comment}
\begin{equation}
 \label{phase_result_1}
  \delta\Phi^{(T^3)}\cong\frac{ma_B}{32\hbar}\left(\frac{\mu_1-\mu_2}{\mu_B}\right)\left(g+\frac{\mu_1+\mu_2}{3\mu_B}\,a_B\right)T^3
\end{equation}
of the interferometer phase with the total time $T\cong 4T_1$. In our experiment $T_d=2.6\,\mu$s, with $T_1$ having a maximal value of $70\,\mu$s.

The maximum contrast displayed by the gray lines is first measured by performing only the spin-echo sequence ($\pi/2-\pi-\pi/2$) without the
magnetic field gradients and changing the phase of the closing RF $\pi/2$-pulse. The maximal visibility is limited by imperfections in the RF
pulses. The echo sequence also allows us to cancel out the contribution to the interferometer phase from the bias magnetic field, and to
increase the coherence time.

As a result, this interferometric technique allows us to measure the magnetic field acceleration, $a_{B}^{fit}=273.16 \pm 0.09$ ${\rm m/s^2}$.
On the other hand, the magnetic field gradient $\partial B_y/\partial z$ was determined independently by the time-of-flight (TOF) technique
where the atomic ensemble is released from the magnetic trap, and a single magnetic field gradient pulse is applied. Measuring the final
position of the atomic ensemble after some TOF, we obtain $a_{B}^{TOF}=271\pm6$ ${\rm m/s^2}$. The difference in measurement errors clearly
shows that our $T^3$-SGI provides a precise measurement of the magnetic field gradient.

Due to the stable current in the external coils, the fluctuations of the homogeneous bias field are relatively small. The phase noise is
mainly proportional to the amplitude of the magnetic field of the gradient pulses originating from the chip currents \cite{Machluf2013}.
Positioning the atoms near the center of the magnetic field quadrupole created solely by the three chip wires, reduces the phase noise
considerably \cite{Margalit2018}. The fluctuations are further reduced due to the fact that the chip wire current is driven by batteries which
supply a stable voltage modulated using a home-made current shutter. Shot-to-shot charge fluctuations are measured to be $\delta Q/Q = 3.6
\cdot 10^{-3}$ where $Q$ is the total charge in a single pulse.

Next we consider the case when $T_1\ll T_d$ and keep the absolute value of the relative momentum $\delta p_0\equiv m a_B T_1
(\mu_1-\mu_2)/\mu_B$ between the two paths constant, that is we take the magnetic field gradient pulses to be a delta function. In this limit
the interferometer phase
\begin{equation}
 \label{phase_result_2}
  \delta\Phi^{(T^2)}\cong\frac{\delta p_0}{4\hbar}\,g T^2
\end{equation}
following from Eq.~(\ref{phase_result}) scales quadratically with the total time $T\cong2T_d$, since we maintain a piece-wise constant
momentum difference between the two arms similar to the $T^2$-SGI \cite{Margalit2018}, or the Kasevich-Chu interferometer
\cite{Zimmermann2019}. However, in the case of the $T^2$-SGI the momentum transfer $\delta p_0$ is provided by the magnetic field gradient
rather than the laser light pulse.

\begin{figure}
\centerline{
\includegraphics[width=\columnwidth]{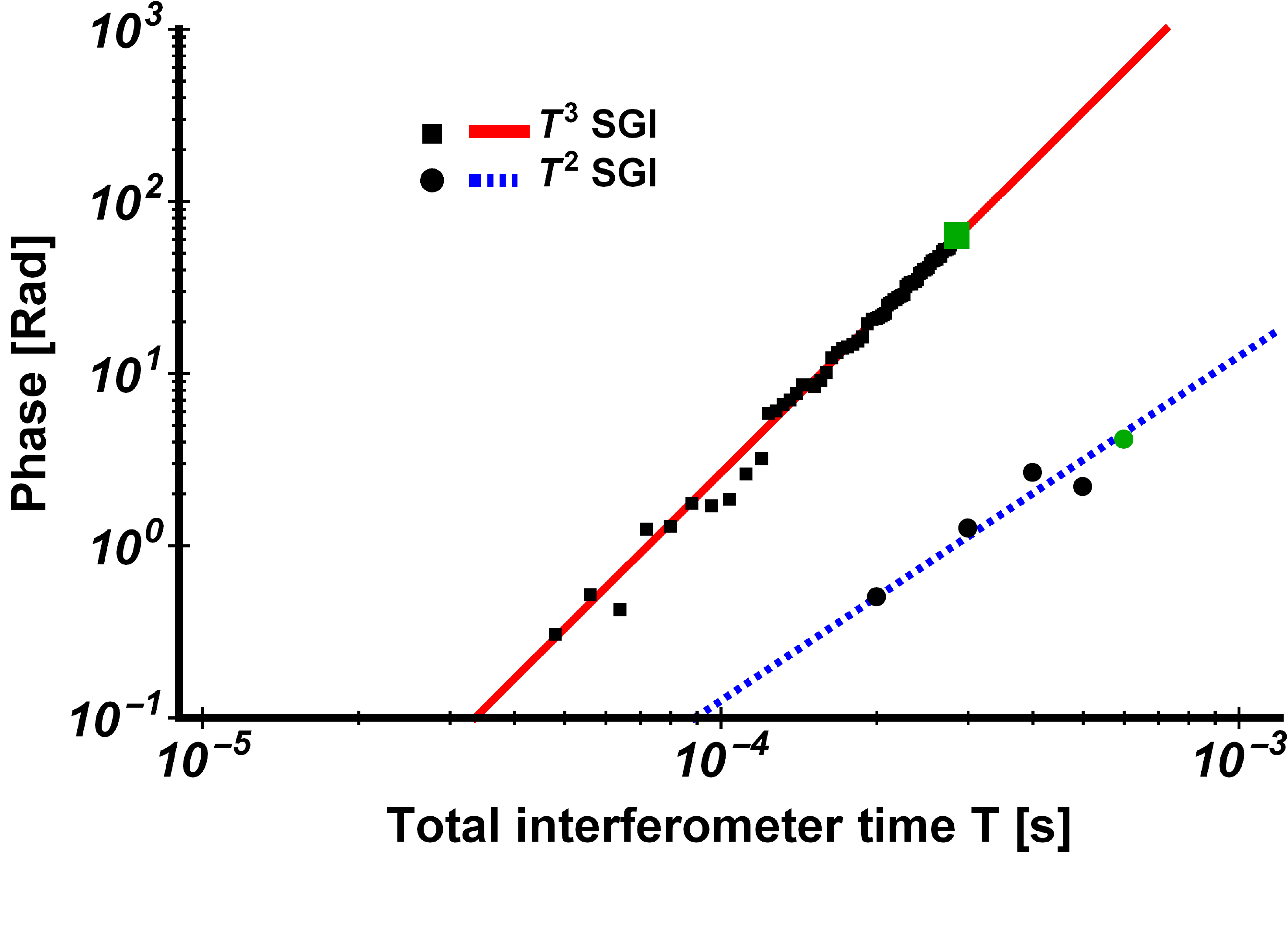}}
\caption{Comparison between the scalings of the interferometer phases $\delta\Phi^{(T^3)}$ (solid red curve) and $\delta\Phi^{(T^2)}$ (dashed
blue curve) given by Eqs. (\ref{phase_result_1}) and (\ref{phase_result_2}), respectively. The squares are data from
Fig.~\ref{fig:closedloop}, while the dots are from our $T^2$-SGI realization \cite{Margalit2018}.
The green square represents the largest accumulated phase of the $T^3$-SGI in its current configuration with $T_{\rm max}=285\,\mu{\rm s}$ at
the point of $\approx 30\%$ contrast. As a reference, the green dot indicates the largest accumulated phase observed with a $T^2$-SGI also at
$\approx 30\%$ contrast. The latter is significantly smaller although the magnetic field gradients and the maximal time $T_{\rm max} =
924\,\mu{\rm s}$ are larger than those of the $T^3$-SGI by a factor of 2.3 and 3.2, respectively \cite{differences}.
}

\label{fig:phase}
\end{figure}

Finally, in Fig.\,\ref{fig:phase} we compare the scaling of the interferometer phases $\delta\Phi^{(T^3)}$ (solid red curve) and
$\delta\Phi^{(T^2)}$ (dashed blue curve) given by Eqs. (\ref{phase_result_1}) and (\ref{phase_result_2}), respectively.
In squares are the data points from Fig.~\ref{fig:closedloop} and in dots are data from our $T^2$-SGI \cite{Margalit2018}.
It is clear that the $T^3$-SGI significantly outperforms the $T^2$-SGI with respect to total phase accumulation.

\noindent{\it Summary.--} We have realized a novel matter-wave interferometer, being unique in several ways: (i) It opens a road towards
testing the Humpty-Dumpty hypothesis, (ii) The phase scales purely as $T^3$, constituting an observation of the Kennard phase, and
(iii) It does not utilize light to create beam splitters and mirrors, enabling a novel probe. A
future challenge will be to find configurations with this, or even an improved, phase scaling, which are more sensitive to external forces.

\begin{acknowledgments}

We thank Zina Binstock for the electronics and the BGU nano-fabrication facility for providing the high-quality chip.
This work is funded in part by the Israel Science Foundation (grant No. 856/18) and the German-Israeli DIP projects
(Hybrid devices: FO 703/2-1, AR 924/1-1, DU 1086/2-1) supported by the DFG. We also acknowledge support from the Israeli Council for Higher
Education (Israel).
M.A.E. is thankful to the Center for Integrated Quantum Science and Technology ($IQ^{ST}$) for its generous financial
support. W.P.S. is grateful to Texas A$\&$M University for a Faculty Fellowship at the Hagler Institute for
Advanced Study at Texas A$\&$M University, and to Texas A$\&$M AgriLife Research for the support of this work.
The research of the $IQ^{ST}$ is financially supported by the Ministry of Science, Research and Arts, Baden-Württemberg.
F.A.N. is grateful for a generous Laboratory University Collaboration Initiative (LUCI) grant from the Office of the Secretary of Defense.
\end{acknowledgments}


\begin{thebibliography}{10}
\bibitem{differences}
The dashed blue curve results from Eq.\,(\ref{phase_result_2}) for $a_B=626$\,m/s$^2$, which is larger than the one used in the current
experiment. For the $T^2$-SGI we present only data for which $T_1\ll T_d$ as required for Eq.\,(\ref{phase_result_2}) to be valid.
\end{thebibliography}
\end{document}